\documentclass[fleqn,usenatbib]{mnras}

\usepackage{newtxtext,newtxmath}
\usepackage[T1]{fontenc}

\DeclareRobustCommand{\VAN}[3]{#2}
\let\VANthebibliography\thebibliography
\def\thebibliography{\DeclareRobustCommand{\VAN}[3]{##3}\VANthebibliography}

\usepackage{graphicx}	% Including figure files
\usepackage{amsmath}	% Advanced maths commands
\usepackage{pythonhighlight}
\usepackage{siunitx}
\usepackage{natbib}
\title[FRB 20201124A classification with unsupervised ML]{Classifying a frequently repeating fast radio burst, FRB 20201124A, with unsupervised machine learning} 

\author[Bo-Han Chen et al.]{
Bo Han Chen$^{1}$\thanks{E-mail:oliver.bo.han.chen@gmail.com},
Tetsuya Hashimoto$^{2}$,
Tomotsugu Goto$^{1}$,
Bjorn Jasper R. Raquel$^{3}$,
\and
Yuri Uno$^{2}$,
Seong Jin Kim$^{1}$,
Tiger Y.-Y. Hsiao$^{4}$,
and Simon C.-C. Ho$^{5}$
\\
$^{1}$Institute of Astronomy, National Tsing Hua University, No. 101, Section 2, Kuang-Fu Road, Hsinchu City 30013, Taiwan\\
$^{2}$Department of Physics, National Chung Hsing University, No. 145, Xingda Rd., South Dist., Taichung, 40227, Taiwan\\
$^{3}$Department of Earth and Space Sciences, Rizal Technological University, Boni Avenue, Mandaluyong, 1550 Metro Manila, Philippines\\
$^{4}$Department of Physics and Astronomy, Johns Hopkins University, Baltimore, MD 21218, USA\\
$^{5}$Research School of Astronomy and Astrophysics, The Australian National University, Canberra, ACT 2611, Australia\\
}

\date{Accepted 2023 March 20. Received 2023 March 17; in original form 2022 October 28}

\pubyear{2022}

\begin{document}
\label{firstpage}
\pagerange{\pageref{firstpage}--\pageref{lastpage}}
\maketitle

\begin{abstract}
Fast radio bursts (FRBs) are astronomical transients with millisecond timescales. Although most of the FRBs are not observed to repeat, a few of them are detected to repeat more than hundreds of times. There exist a large variety of physical properties among these bursts, suggesting heterogeneous mechanisms of FRBs.
In this paper, we conduct a categorisation on the extremely frequently repeating FRB 20201124A with the assistance of machine learning, as such techniques have the potential to use subtle differences and correlations that humans are unaware of to better classify bursts. The research is carried out by applying the unsupervised Uniform Manifold Approximation and Projection (UMAP) model on the FRB 20201124A data provided by Five-hundred-meter Aperture Spherical radio Telescope (FAST).
The algorithm eventually categorises the bursts into three clusters. In addition to the two categories in previous work based on waiting time, a new way for categorisation has been found.
The three clusters are either high energy, high frequency, or low frequency, reflecting the distribution of FRB energy and frequency. 
Importantly, a similar machine learning result is found in another frequently repeating FRB20121102A, implying a common mechanism among this kind of FRB. This work is one of the first steps towards the systematical categorisation of the extremely frequently repeating FRBs.
\end{abstract}


%%%%%%%%%%%%%%%%%%%%%%%%%%%%%%%%%%%%%%%%%%%%%%%%%%

%%%%%%%%%%%%%%%%% BODY OF PAPER %%%%%%%%%%%%%%%%%%

\section{Introduction}

Fast radio bursts (FRBs) are astronomical transients that show sudden brightening at radio wavelengths (e.g. \citealt{Lorimer_2007}).  Most of the FRBs happen in milliseconds timescale \citep[e.g.][]{Petroff2016} and occur at cosmological distances. FRBs are usually classified as either non-repeating or repeating, where they are defined as one-off bursts and repeated bursts detected from single FRB sources, respectively. Numerous FRB theories have been proposed \citep[e.g.][]{Platts2019} to account for the physical origin of either non-repeating or repeating FRBs. The proposed hypotheses include neutron star mergers \citep{Yamasaki2017}, pulsar–black hole interactions \citep{Bhattacharyya2017}, active galactic nuclei activities \citep{Vieyro2017AMF}, etc. In spite of the intensive observations and modelling, so far, the origin of FRBs remains a mystery.

Most of the repeating FRB sources only have a handful of observed burst events (\citealt{amiri2021first}). However, a highly active repeater could have more than 1000 bursts detected.  The well-known FRB20121102 \citep{Li2021} and FRB 20201124A \citep{Xu2021} are both examples of a highly active repeater also known as frequently repeating FRBs. This kind of FRBs are rarely seen, but the multitude of their bursts allows us to perform categorisation with the aid of machine learning. 

 FRB 20201124A is the most active FRB known so far \citep{Xu2021}. A radio observational campaign of Five-hundred-meter Aperture Spherical radio Telescope (FAST) is dedicated to monitoring FRB 20201124A which provided the necessary data for this research. The campaign monitored FRB 20201124A from April 1, 2021 up until June 11, 2021, which cumulatively has 96.9 hours of observation time. The released catalogue contains 1863 bursts, and each burst is characterised by at most 27 physical properties \citep{Xu2021}.

These 1863 bursts are detected with a signal-to-noise ratio (S/N) > 7, with 913 bright bursts reaching S/N > 50. The burst rate of FRB 20201124A evolves slowly between $5.6\ \si{hr^{-1}}$ to $45.8\ \si{hr^{-1}}$. The burst flux density ranges from 0.005 Jy to 11.5 Jy, and the inferred isotropic luminosity spans from $5 \times 10^{37}\ \si{erg\ s^{-1}}$ to $3 \times 10^{40}\ \si{erg\ s^{-1}}$ \citep{Xu2021}. The data provided by the observation is homogeneous, therefore analysing it with unsupervised machine learning is a feasible direction.

There are already several works that applied machine learning to study FRBs. Most recently, \citealt{Hewitt2021} used a supervised deep-learning algorithm to classify FRB candidates from the FRB121102 data which is detected by the Arecibo Telescope. Before that \cite{Agarwal2020} also presented a classification for FRB candidates from Australian Square Kilometre Array Pathfinder (ASKAP) detection using deep neural networks. In an earlier work, \cite{Wagstaff_2016} trained and deployed a machine learning classifier that marks each detection as either a known pulsar, artefact due to interference, or potential new FRB discovery. Suggesting that unsupervised machine learning is a promising way to identify misclassified FRB repeaters. For example, \cite{Chen2021} successfully uncloaked the hidden repeating FRB candidates from non-repeating FRBs in the CHIME/FRB observation with UMAP, an unsupervised machine learning model. Thus, motivating subsequent FRB studies such as \citealt{SeongJin2022} and \citealt{Tetsuya2022}.

In this paper, an unsupervised dimensionality reduction algorithm is used and followed by an unsupervised clustering algorithm. The dimensionality reduction algorithm we utilised is known as Uniform Manifold Approximation and Projection (UMAP), a technique for dimension reduction based on manifold learning techniques and ideas from topological data analysis \citep{mcinnes2018umap}. The UMAP algorithm maps the observational parameters of each FRB to a 2D embedding plane after training on the features of the training samples. This projection aggregates the FRB samples having similar physical properties and repels those that do not. As a result, the algorithm provides us with a way to systematically classify the bursts of FRB 20201124A. Next, an unsupervised clustering algorithm known as K-means is used to label the clusters resulting from the projection.

In conclusion, we find that the bursts in FRB 20201124A could be roughly classified into 3 clusters, rather than the previously known two \citep{Xu2021}. The classification is mainly dominated by the energy and frequency parameter of the bursts, while both the arrival time and the waiting time do not have significant effect on the clustering result. These findings are different from the bimodal distribution of waiting time proposed in \cite{Xu2021}. This work is a first step towards the systematic categorisation of the extremely frequently repeating FRBs.

This work is organised as follows. We describe our data composition and model configuration in Section~\ref{sec:Data}. Our UMAP Model classification result is described in Section~\ref{sec:Result}. We present the Discussion in Section~\ref{sec:Discussion}, followed by Conclusions in Section~\ref{sec:Conclusions}. 

\section{Data composition and model configuration}
\label{sec:Data}

\subsection{Sample and Data Selection}
\label{sec:Sample and Data Selection}

The data for FRB 20201124A in this work is provided by the observational campaign of FAST \citep{Xu2021}. The catalogue records 1863 FRBs at a frequency range between 1.0 GHz to 1.5 GHz from April 1, 2021 up until June 11, 2021. There are 27 observational parameters available in the catalogue. We included 8 parameters in the main unsupervised machine learning experiment, where all bursts must have the measurements of these 8 features. 118 samples are excluded due to the missing parameter values, thus finally 1745 FRB are utilised in this paper.  We further discuss the details of each observational parameter in Section~\ref{sec:The observational parameters}.

\subsubsection{The observational parameters}
\label{sec:The observational parameters}

A total of 8 observational parameters are included in our unsupervised training.  All of these parameters come from the original FRB 20201124A data provided by FAST. Brief descriptions of the parameters are as follows \citep[see][for details]{Xu2021}.

\begin{itemize}

\item \textbf{Barycentrical arrival time (BAT) } : The Barycentrical arrival time of the FRBs in units of Modified Julian Date (MJD), measured from the centroid of the best-matched boxcar filter. The value of this parameter is ranged from 59307.33458 to 59360.17901.

\item \textbf{S/N} : The signal-to-noise ratio of the FRBs. The value of this parameter is ranged from 7.14 to 2885.83. 

\item \textbf{Peak flux density} $\mathrm{(Jy)}$ : The peak flux of the FRBs. The value of this parameter is ranged from 0.005 to 11.505.

\item \textbf{Fluence} $\mathrm{(Jy}\cdot\mathrm{ms)}$ : The integration of the pulse flux density with respect to the time of the FRBs. The value of this parameter is ranged from 0.0173 to 67.3181.

\item \textbf{Equivalent width} $\mathrm{(ms)}$ : A measure of the time width of FRBs, defined by the fluence divided by the pulse peak flux density. The value of this parameter is ranged from 1.556 to 28.36.

\item \textbf{Bandwidth} $\mathrm{(MHz)}$ : The bandwidth of the FRBs. The value of this parameter is ranged from 30.151 to 485.84.

\item \textbf{Peak frequency} $\mathrm{(MHz)}$ : The peak frequency of the FRBs. The value of this parameter is ranged from 1000 to 1500.

\item\textbf{Waiting time} $\mathrm{(s)}$ : The time difference between each FRB and its next one. We impose the log scale on the value for training since it spans several orders of magnitude.  We removed 118 FRB samples at the end of each observation session because their waiting time cannot be measured. The value of this parameter ranges from 0.004 to 1325.108 seconds.

It is worth noting that, as the FRBs in this paper originated from a single source, the cosmological effect is not opposing discrepant effect on them. As a result, we don't consider the cosmological effect correction in this paper.

\end{itemize}

\subsubsection{The Statistical Information Regarding the Parameters}
\label{sec:The statistical information regarding the parameters}

\begin{figure*}
	\includegraphics[width=0.8\textwidth]{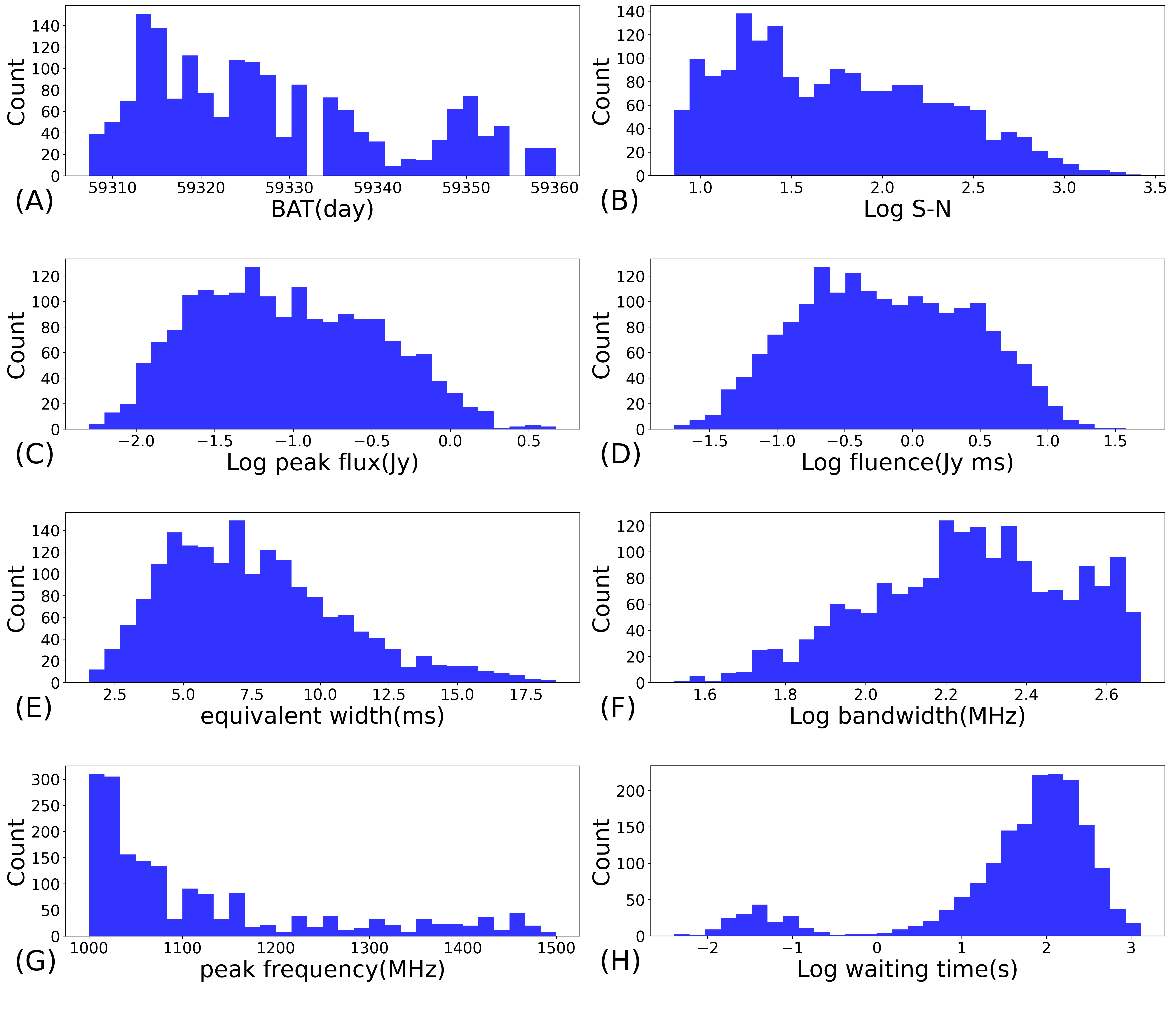}
    \caption{This figure shows the observational parameter distributions, which are the input data we apply for our unsupervised learning. See Section~\ref{sec:The observational parameters} for the details.}
    % remove the fig title
    \label{fig:Fig1}
\end{figure*}

Our data is comprised of the 1745 FRBs originating from FRB 20201124A, and this sample is the foothold of our unsupervised machine learning model. In order to further understand the basic composition of our research, we plot the distribution of the parameters mentioned in Section~\ref{sec:The observational parameters}, shown in Fig.~\ref{fig:Fig1}.

\subsection{Unsupervised Machine Learning Model and Configuration}

We utilised 1745 FRBs provided by the FAST observation campaign, and there are 8 physical parameters for each FRB. Serving as our input data, we managed to conduct two unsupervised machine learning analyses on this data. The applied machine learning techniques are UMAP, an algorithm based on topological data analysis and manifold learning, and K-means, an algorithm for unsupervised clustering.

The training of UMAP involves two stages.  In the first stage, the algorithm constructs the local connectivity of the data manifold by computing the distances of each data point's $k$ nearest neighbours under a relative scale of the distance to nearest neighbours.  As a result, $k$ is one of the UMAP hyperparameters, referred to as \pyth{n_neighbors}.  \pyth{n_neighbors} manipulates the model's balance between the local and the global manifold structure of the data.  A large \pyth{n_neighbors} demands UMAP to consider a large cluster of data points when computing the local connectivity.  On the other hand, a small \pyth{n_neighbors} allows the algorithm to concentrate on the local structure of the data. We discuss our choice of \pyth{n_neighbors} in Sec.~\ref{sec:Result}.

The second stage of UMAP is to map the FRBs (represented by 8 parameters) to low-dimensional representations. Thus, we need another hyperparameter \pyth{n_components} to decide the resulting dimensionality of the reduced dimension.  In our work, we adopt \pyth{n_components} = $2$, which means we project the physical properties of FRBs onto a 2D plane.

To find a 2-dimensional representation that matches the topological structure of the 8-dimensional data, UMAP performs a stochastic gradient descent by a specific cross-entropy function. The gradient descent provides an attractive force between the points where the scaled distance mentioned above is short in high dimensional space and provides a repulsive force between the points whenever the distance is large.  In order to prevent the resulting low dimensional projection from clumping together, another hyperparameter \pyth{min_dist} is induced to constrain the minimum Euclidean distance between the projected points.  However, this hyperparameter is not affecting our result significantly. In this paper, \pyth{min_dist} is set to have the default value of $0.1$.

The result provided by UMAP is followed by the unsupervised clustering method K-means. It is used to group data into clusters based on similarity. It is one of the most widely used clustering algorithms and is relatively simple to implement. The algorithm works by first initializing a set of cluster centres, then iteratively assigning data points to the closest cluster centre, and updating the cluster centres to be the mean of the points assigned to that cluster. This process is repeated until the cluster centres stop changing or a maximum number of iterations is reached. The main advantage of the K-means algorithm is its speed and simplicity, but it requires a preset cluster number \pyth{n_cluster} to determine the total cluster numbers. We discuss our choice of \pyth{n_cluster} in Sec.~\ref{sec:Result} and Appendix~\ref{sec:Additional clustering result}.

This section concludes with a simple introduction to the UMAP algorithm and the hyperparameter setting we use.  Readers can refer to \citet{mcinnes2018umap} for the mathematical background of UMAP.

\section{Result}
\label{sec:Result}

\subsection{UMAP Model Classification Result}

\begin{figure}
    \centering
    \includegraphics[width=\columnwidth]{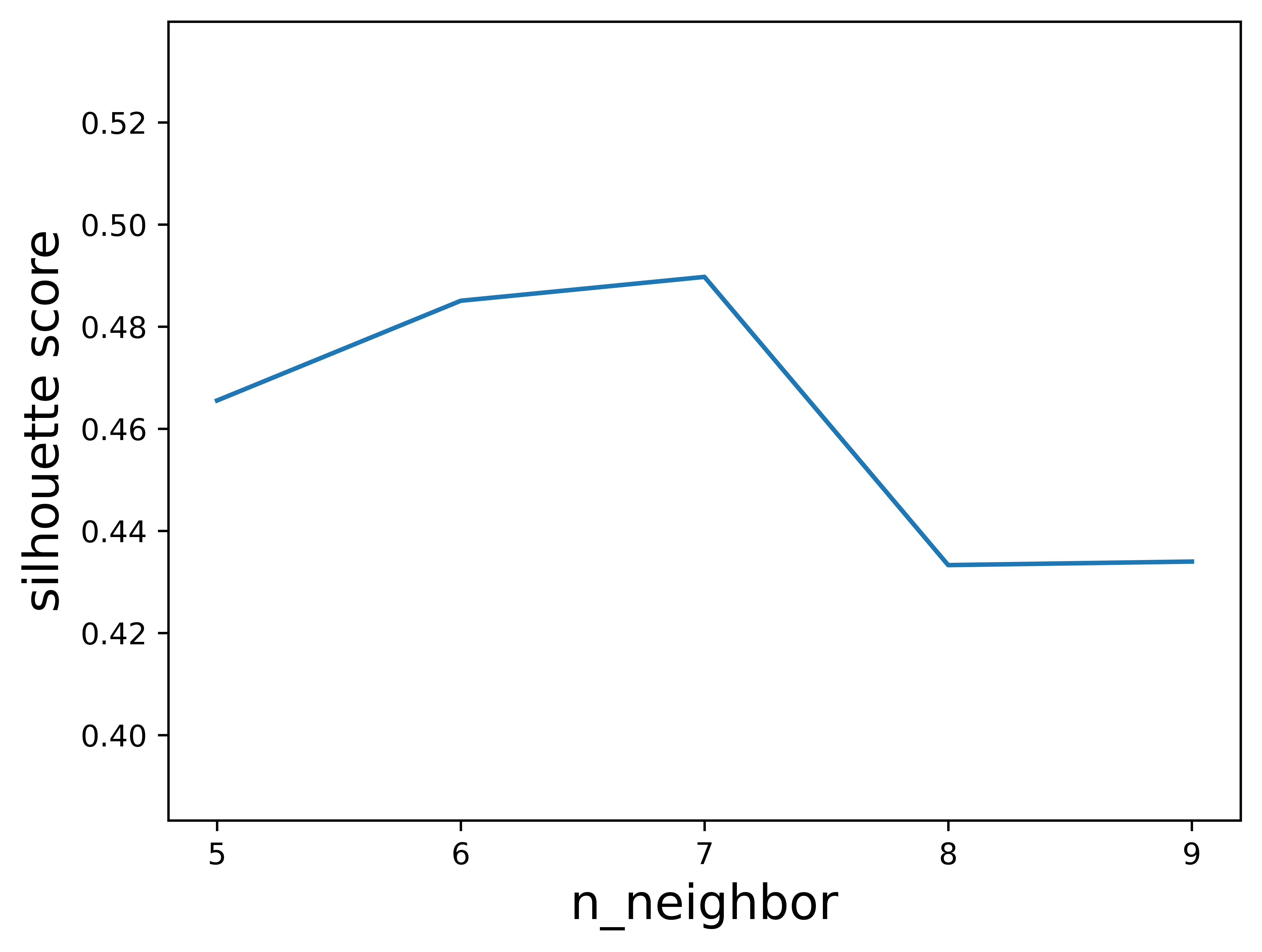}
    \caption{The silhouette score result from each \pyth{n_neighbor} setting for the UMAP algorithm.}
    \label{fig:n_neighbors}
\end{figure}

Our unsupervised UMAP model makes the projection and then K-means identify the clusters. We expect the projection of the FRBs to have distinct clusters on the embedding plane. In order to do that, we need to properly set up the UMAP hyperparameter \pyth{n_neighbors}. Thus, we evaluate the silhouette score \citep{ROUSSEEUW198753} result from each \pyth{n_neighbors}, presented in Fig.~\ref{fig:n_neighbors}. The silhouette score is a metric for evaluating the performance of a clustering algorithm. The silhouette score ranges from -1 to 1, with a score of 1 indicating that all the samples are perfectly matched to their own clusters, and a score of -1 indicating that the samples are poorly matched to their own clusters. The highest silhouette score is obtained when \pyth{n_neighbors} = 7. Therefore, we use \pyth{n_neighbors} = 7 in the paper. 

\begin{figure}
    \centering
    \includegraphics[width=\columnwidth]{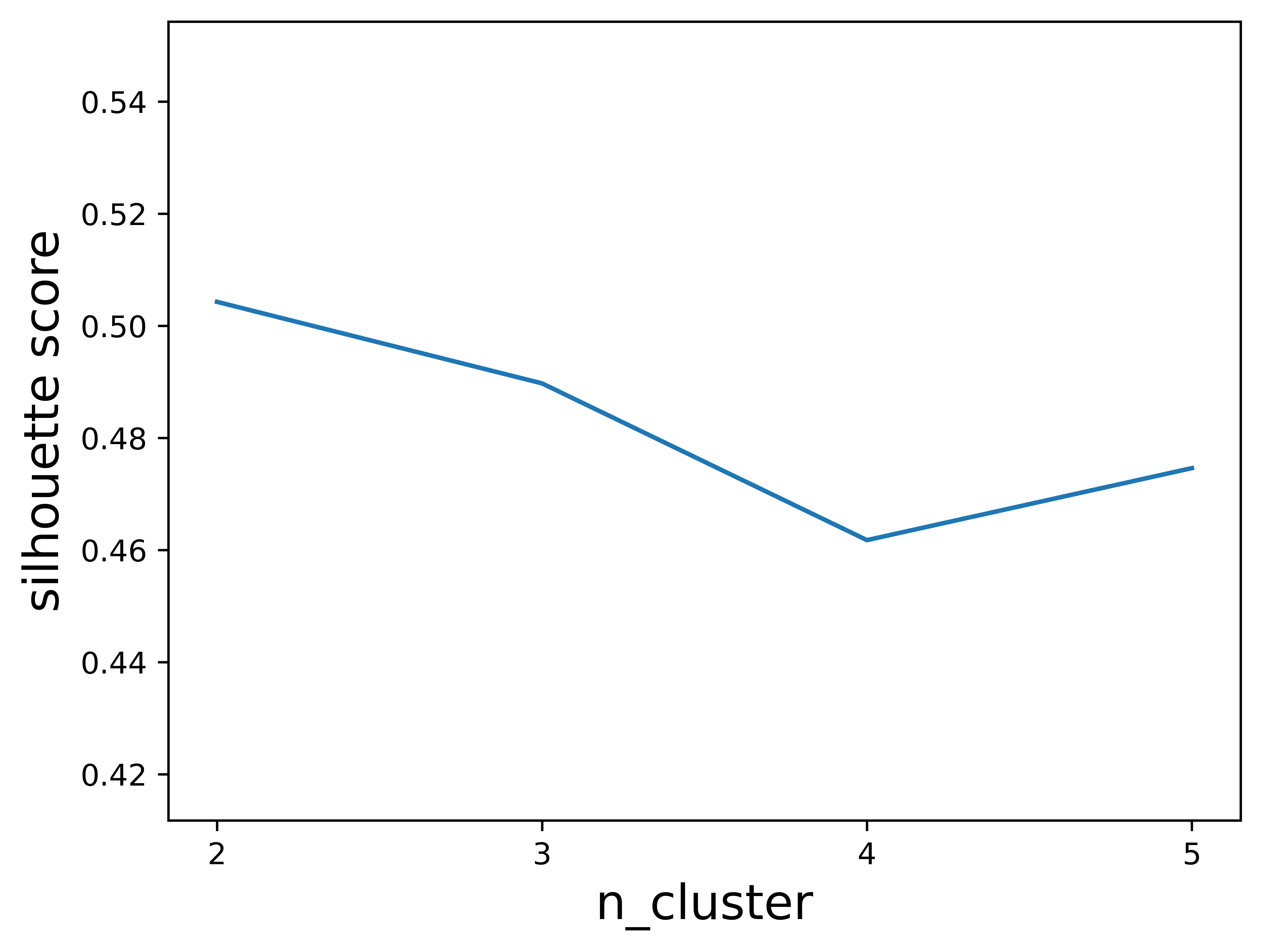}
    \caption{The silhouette score result from each \pyth{n_cluster} setting for the K-means clustering algorithm.}
    \label{fig:n_cluster}
\end{figure}

K-means clustering algorithm requires a predetermined value for the hyperparameter \pyth{n_cluster} to decide the number of the resulting clusters. In order to select the best value, we evaluate the silhouette score of each \pyth{n_cluster} setting, presented in Fig.~\ref{fig:n_cluster}. In the figure, we see the clustering has the best silhouette score in \pyth{n_cluster} = 2. However, the embedding plane obviously has more than two clusters. Therefore, we exclude \pyth{n_cluster} = 2 from further analysis. Except for \pyth{n_cluster} = 2, \pyth{n_cluster} = 3 provides the best silhouette score.
 Therefore, we apply \pyth{n_cluster} = 3, separating the embedding plane into 3 areas.

\begin{figure}
    \centering
    \includegraphics[width=\columnwidth]{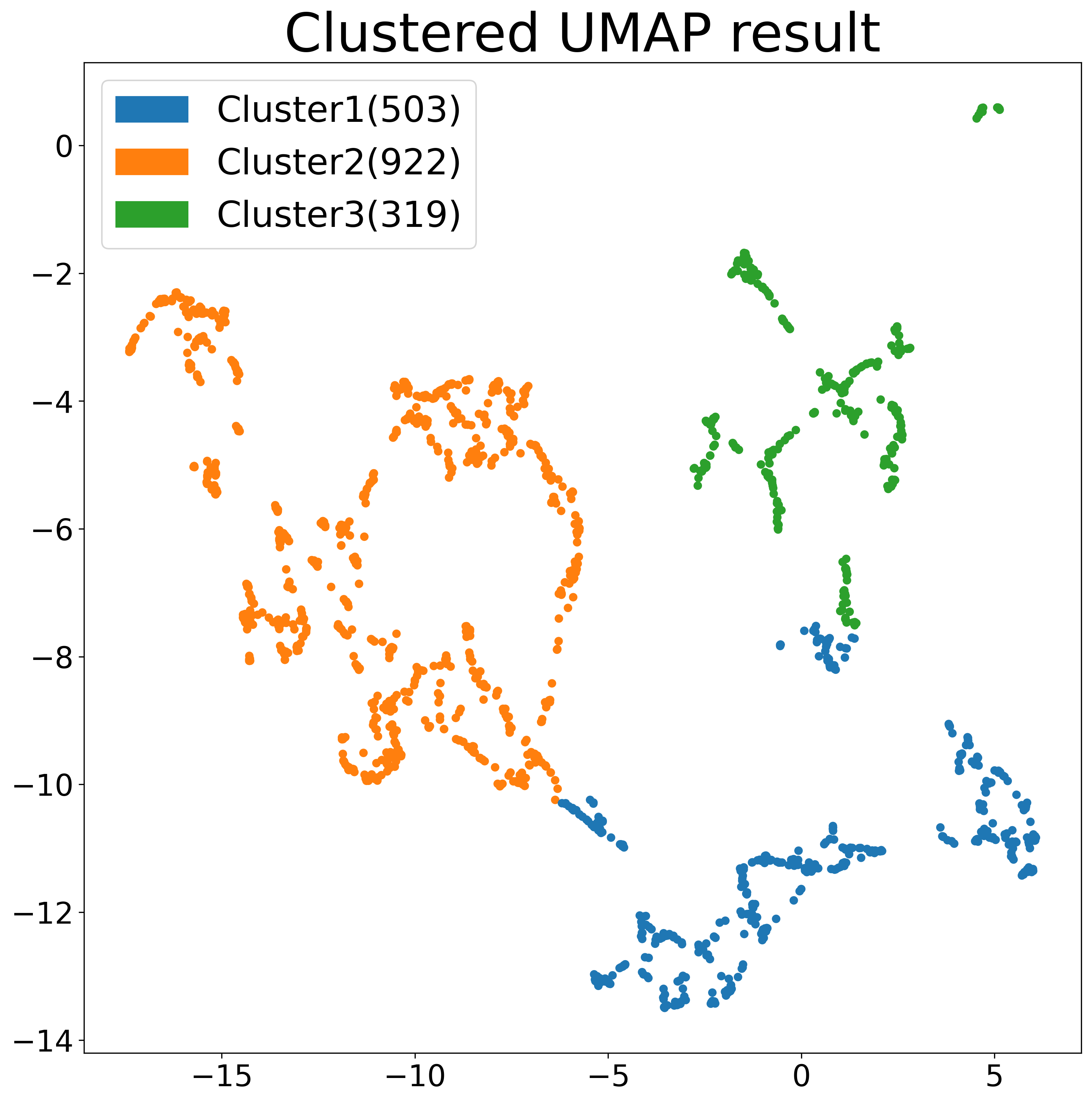}
    \caption{The unsupervised machine learning result of the FRB 20201124A. The distances among the points represent the similarities of their physical parameters, but the $x$ and $y$ axes do not have explicit physical implications. 1745 FRBs are projected to the embedding plane according to their physical properties. Classification is carried out with the K-means clustering algorithm, and the resulting clusters are marked in different colours.}
    \label{fig:Fig2}
\end{figure}

We present the result of our unsupervised machine learning in Fig.~\ref{fig:Fig2}. Aggregation of the FRBs on the embedding plane implies a similar physical property. The data points on the plane have been classified by the K-means clustering algorithm. We denote the three clusters as Cluster 1, Cluster 2 and Cluster 3. Cluster 2 is large and sparse, including more than half of the samples. In contrast, the structure of cluster 1 and cluster 3 are denser. This fact implies the clusters on the right-hand side have similar physical parameters. The visual inspection suggests that Cluster 1 could be composed of two sub-clusters. However, due to the restrictions of the clustering algorithm, we are not able to separate these two sub-clusters.

The clusters presented on the embedding plane suggest that there is indeed heterogeneity in the behaviour of FRB 20201124A. As a result, we look into the physical property distribution on the embedding plane in the following section and figure out which factors contribute to the formation of these clusters.

\subsection{Physical Property Distribution of Each cluster}
\begin{figure*}
    \centering
    \includegraphics[width=0.7\textwidth]{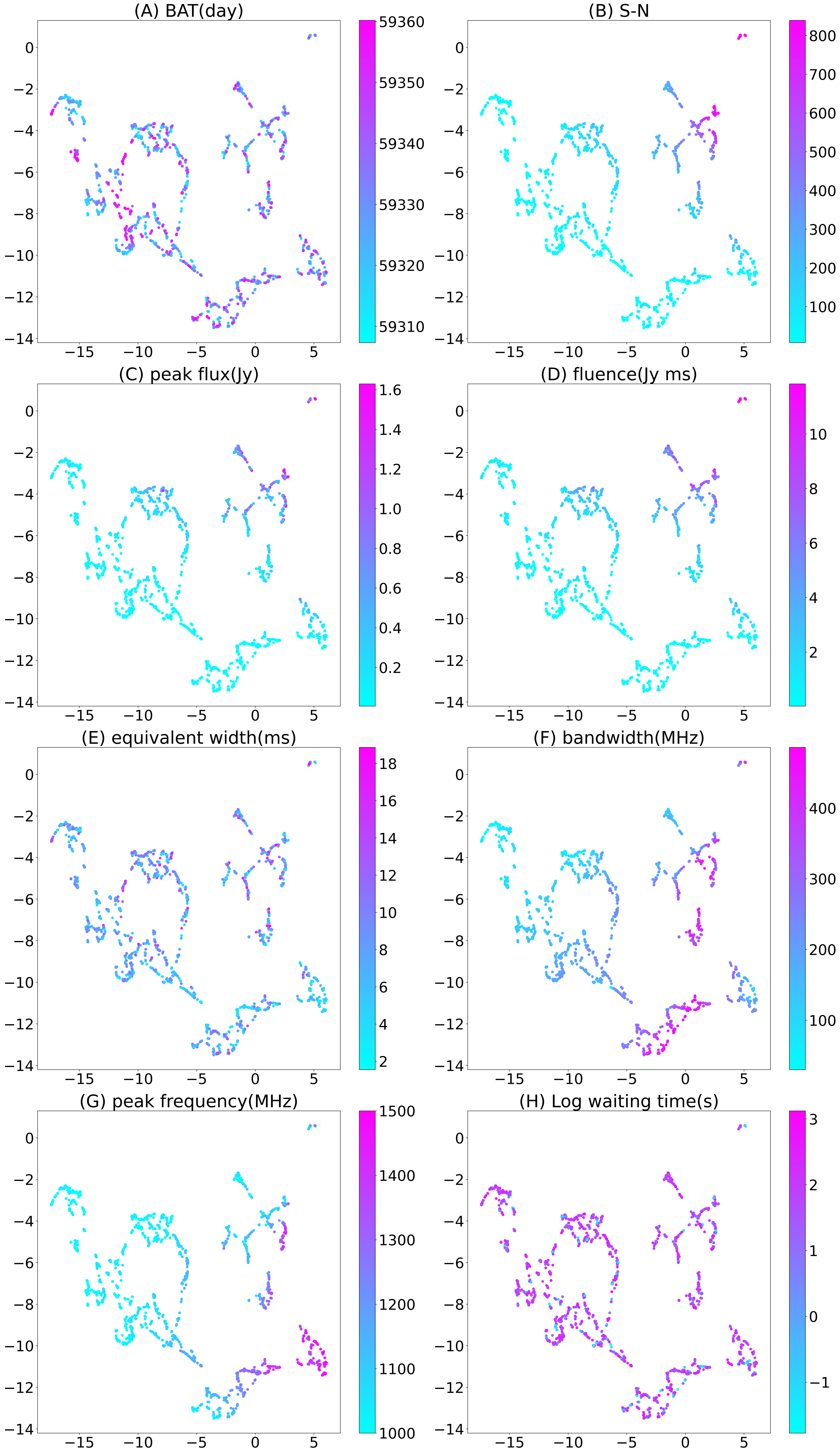}
    \caption{The value distribution of each FRB on the embedding plane.  The numerical scale is the same as Fig. 4. The range of the color bar in each panel is limited to $\pm 3$ standard deviations from the average for clear visualization.}
    \label{fig:Fig3}
\end{figure*}

We colour the embedding plane according to the values of the observational parameters as shown in Fig.~\ref{fig:Fig3}. In the figure, we see S/N, peak flux, fluence, bandwidth and peak frequency present clear correlation with each other, implying these are the factors that significantly influenced the machine learning results. The distributions of BAT, waiting time and equivalent width are almost random, indicating these parameters have very little to no effect on the result. 

According to Fig.~\ref{fig:Fig3}, the parameters can be classified into 3 groups: 

\begin{itemize}

\item Parameter group A: S/N, peak flux and fluence

\item Parameter group B: bandwidth and peak frequency

\item Parameter group C: BAT, waiting time and equivalent width

\end{itemize}

Parameter group A are the factors highlighting cluster 3. Due to these parameters, bursts in cluster 3 generally have higher S/N, peak flux and fluence. These factors are energy-related. Thus, cluster 3 is the high-energy cluster. Parameter group B accentuates the wavelength property of the FRBs. cluster 1 has high bandwidth and peak frequency, while cluster 2 is exactly the opposite. Intriguingly, cluster 3 includes FRBs with both low and high values of bandwidth and peak frequency. Lastly, parameter group C consists of the parameters not influential for the training. BAT, waiting time and equivalent width are the time factors, suggesting our dimensionality reduction result almost is time-independent.

\section{Discussion}
\label{sec:Discussion}

\subsection{Comparing with the UMAP result on multi-FRB source (CHIME/FRB) catalogue.}
\label{sec:Comparing with the UMAP result on multi-FRB source catalogue}

The analysis in this paper mainly focuses on the 1745 FRBs of FRB 20201124A which are bursts that originates from a single repeating FRB source. In this section, we compare it to the analysis result based on the FRB catalogue with mostly different sources. The catalogue we referred to here was released by Canadian Hydrogen Intensity Mapping Experiment Fast Radio Burst (CHIME/FRB) Project. It includes 536 FRB sources at a frequency range between 400 and 800 MHz from  July 25, 2018 to July 1, 2019 (\citealt{amiri2021first}).

\begin{figure}
    \centering
    \includegraphics[width=\columnwidth]{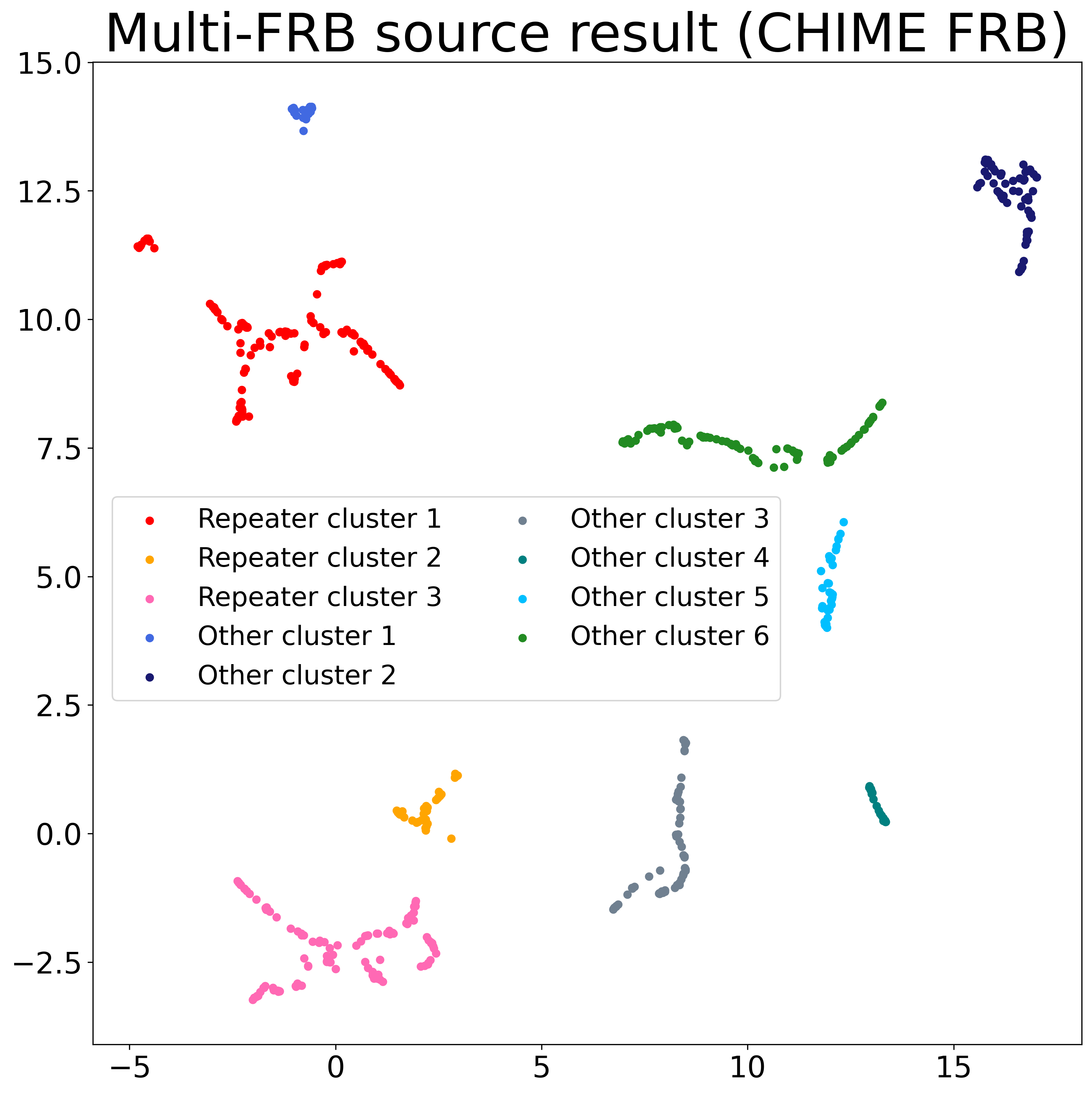}
    \caption{The UMAP analysis based on the CHIME/FRB catalogue, in which FRBs are originated from mostly different sources.}
    \label{fig:Fig4}
\end{figure}

The analysis done for the CHIME catalogue is also based on UMAP, which is originally presented in the work of \cite{Chen2021}. Their work identified the hidden repeating FRBs from the currently one-off FRBs. We transplant the analysis and present it in Fig.~\ref{fig:Fig4}. In order to provide a straightforward comparison, the x and y-axes of the two figures  (Fig. \ref{fig:Fig2} and Fig. \ref{fig:Fig4}) are presented on the same scale. In comparison with our result in Fig.~\ref{fig:Fig2}, the clusters in Fig.~\ref{fig:Fig4} are more distinct and dense. 

The FRBs in Fig.~\ref{fig:Fig2} all originated from one single source, while Fig.~\ref{fig:Fig4} are the FRBs aggregated from 535 sources. As a result, we speculate the different clumping level comes from the different levels of heterogeneity of the underlying data. When the data includes only one single source, the data points are more similar thus the clusters are less distinct. In contrast, when the FRB sources are diverse, the UMAP model can efficiently divide the significantly different FRBs and group FRBs that has some degree of similarity.

In summary, from this comparison, we can infer that the heterogeneity among the FRBs of different sources is higher than that of a single source. The FRBs within FRB 20201124A have different patterns, but the prominence of the pattern difference is definitely lower than the one from the multi-source catalogue.

\subsection{Comparing with the UMAP result on another frequently repeating FRB}
\label{sec:Comparing with the UMAP result on another frequently repeating FRB}

\begin{figure*}
    \centering
    \includegraphics[width=\textwidth]{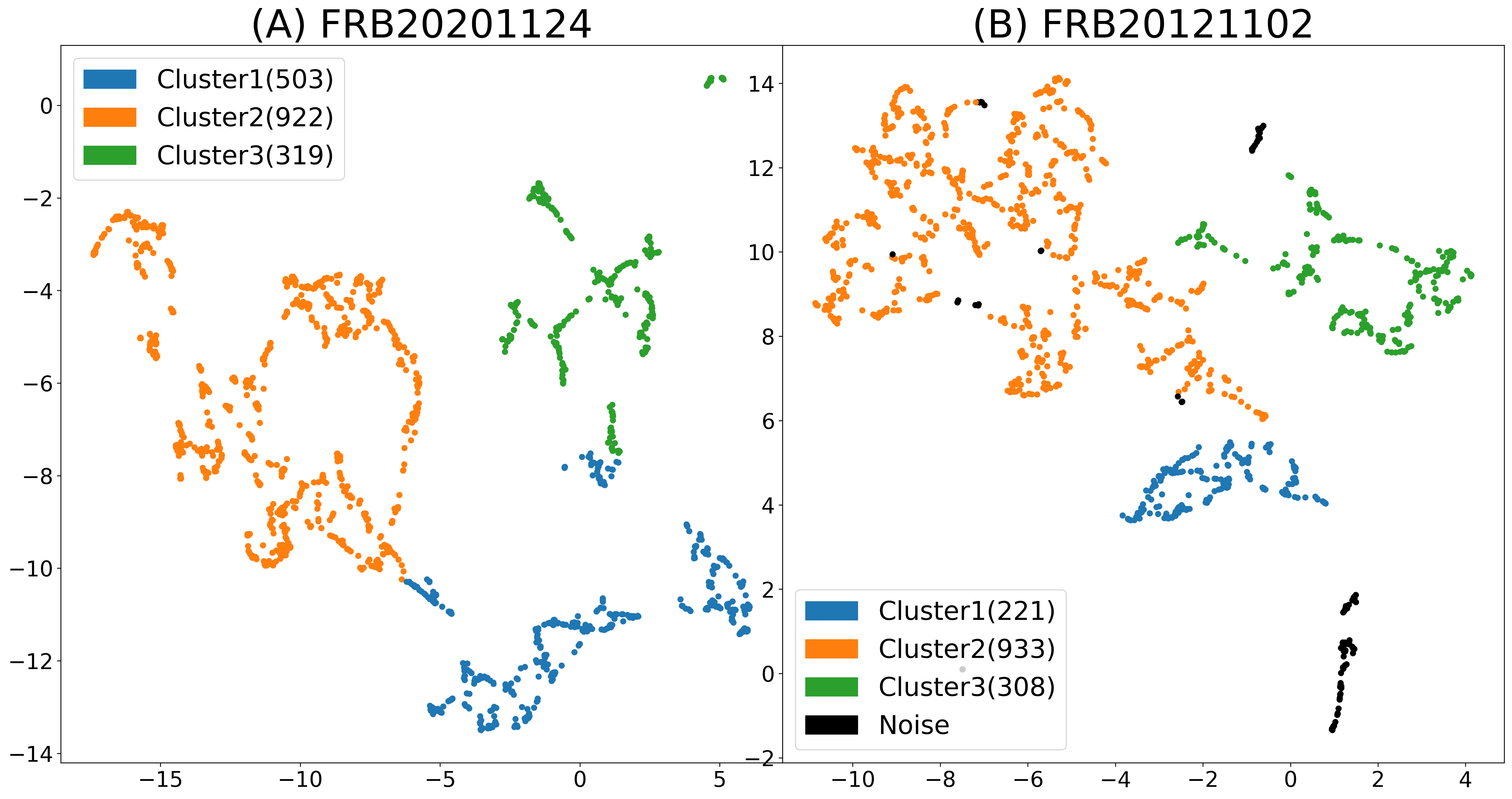}
    \caption{The UMAP result comparison between FRB 20201124A and FRB20121102. The x and y-axes of the two figures are presented on a similar scale.}
    \label{fig:Fig5}
\end{figure*}

FRB 20201124A is not the only observed frequently repeating FRB. FRB20121102 \citep{Li2021} is a source detected to have more than 1000 bursts as well. Raquel et al. (2022 submitted) carried out the UMAP analysis for FRB20121102, which is quite similar to this work. In their work, they included Barycentrical arrival time, dispersion measure, time width, bandwidth, peak flux density, fluence, and energy of the FRBs as the input parameters. Comparing our work, their analysis is presented in Fig.~\ref{fig:Fig5}. Their result is produced by setting the hyperparameter values for \pyth{n_neighbors} $= 6$ and \pyth{min_dist} $= 0$ and clustered by the HDBSCAN algorithm. In this section, we are comparing the best UMAP result of our source and theirs. Since the source data are different, their hyperparameter setting is different from ours. Instead of doing a comparison on identical hyperparameter settings and clustering algorithms, it would be more common in the machine learning community to compare two results in their respective best scenarios (e.g. \citealt{agent57}).

We see that despite the source is different, the UMAP analysis produced results which are similar in nature. Both of them have three main clusters, appearing in 1 large, sparse cluster and 2 small, dense clusters. Moreover, in the lower right corner of both panels, there is a smaller cluster not specified by the clustering algorithm. In summary, the two results are highly matched. The embedding planes reflect the similarity of the FRB pattern. As a result, this fact could be evidence that the frequently repeating FRBs are powered by a similar mechanism.

\section{Conclusions}
\label{sec:Conclusions}

In this work, we applied unsupervised machine learning analysis to the 1745 bursts of FRB 20201124A. For each burst event, we included Barycentrical arrival time, signal-to-noise ratio, peak flux, fluence, equivalent width, and bandwidth of signal as the input data for the algorithm which then projects the data points onto a 2D plane based on their similarity. Our work resulted in three main conclusions:

(i) Our UMAP model divides the FRBs into 3 clusters. The classification is based on either energy or frequency factors. The 3 clusters are either high energy, high frequency, or low frequency. Interestingly, time-related factors did not have a significant effect on our results.

(ii) We compare our result with the work of  \citep{Chen2021} which also has UMAP result based on multi-FRB source data. The multi-FRB source data result has a clumpier, distinct embedding plane. We infer that the heterogeneity among the FRBs from different sources is higher than that of a single source.

(iii) We compare our result with the UMAP result based on another frequently repeating FRB, FRB20121102 (Raquel et al. 2022 submitted). Impressively, their result is very similar to our result, including the scale of the clusters and the relative position of the clusters. This fact suggests that these two frequently repeating FRBs are powered by similar mechanisms.

\section*{Acknowledgements}

%TG and TH acknowledge the supports from the Ministry of Science and Technology of Taiwan through grants 108-2628-M-007-004-MY3 and 110-2112-M-005-013-MY3, respectively.
We sincerely thank the anonymous referee for many insightful comments, which improved the paper significantly.
We sincerely thank Dr H. Xu (Kavli Institute for Astronomy and Astrophysics, Peking University) for assisting with data collection. 
TG acknowledges the support of the National Science and Technology Council of Taiwan through grants 108-2628-M-007-004-MY3 and 111-2123-M-001-008-.
TH acknowledges the support of the National Science and Technology Council of Taiwan through grants 110-2112-M-005-013-MY3, 110-2112-M-007-034-, and 111-2123-M-001-008-.
This work used high-performance computing facilities operated by the
Center for Informatics and Computation in Astronomy (CICA) at National
Tsing Hua University. This equipment was funded by the Ministry of
Education of Taiwan, the National Science and Technology Council of 
Taiwan, and National Tsing Hua University.

%%%%%%%%%%%%%%%%%%%%%%%%%%%%%%%%%%%%%%%%%%%%%%%%%%
\section*{Data Availability}

The utilised data and the UMAP analysis result are available in the article's online supplementary material.



\bibliographystyle{mnras}
\bibliography{mnras_main} % if your bibtex file is called example.bib

%\begin{thebibliography}{99}
%\end{thebibliography}

%%%%%%%%%%%%%%%%%%%%%%%%%%%%%%%%%%%%%%%%%%%%%%%%%%

% %%%%%%%%%%%%%%%%% APPENDICES %%%%%%%%%%%%%%%%%%%%%

\appendix

\section{Additional clustering result}
\label{sec:Additional clustering result}

\begin{figure}
    \centering
    \includegraphics[width=\columnwidth]{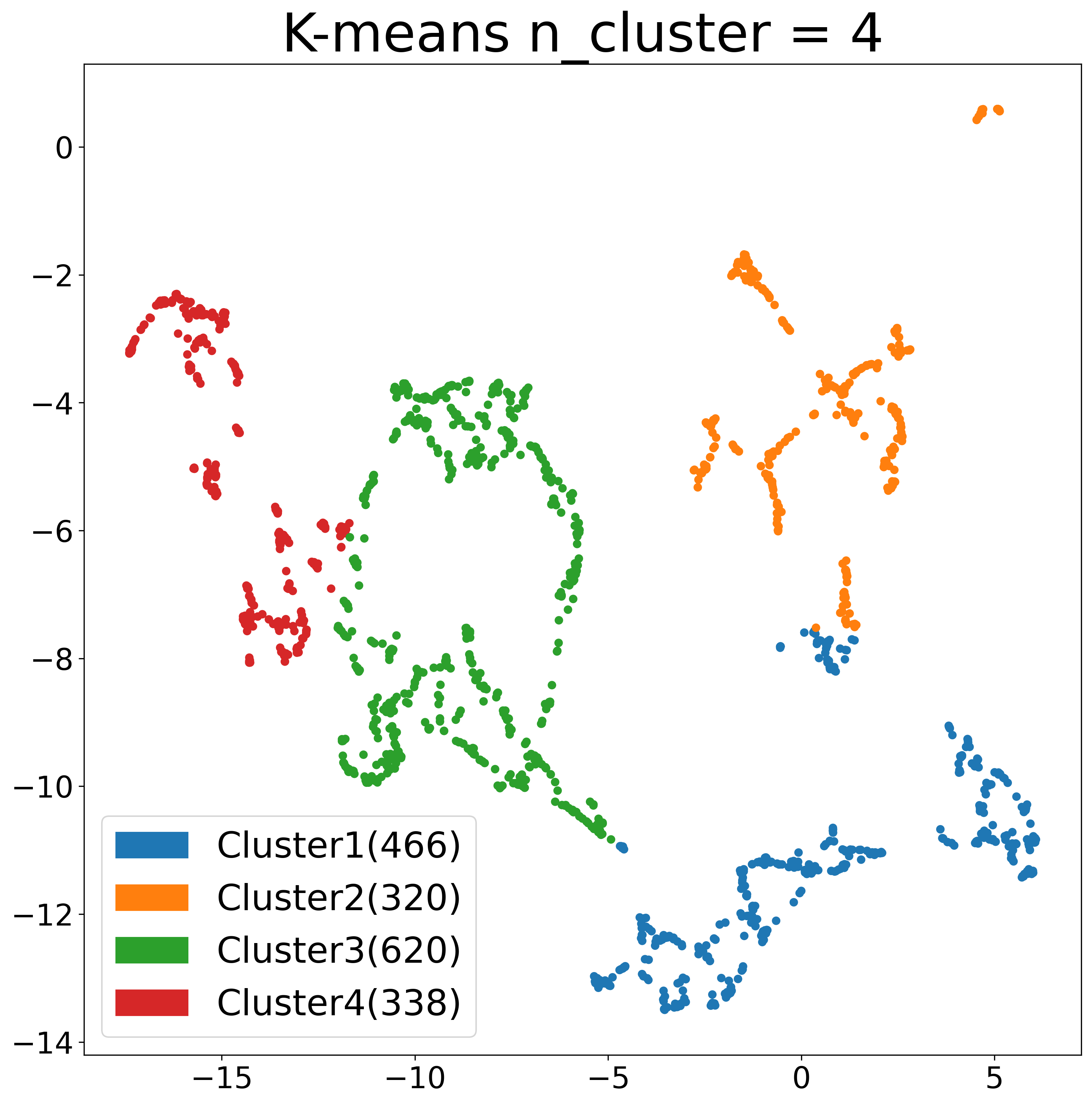}
    \caption{K-means clustering result with \pyth{n_cluster} = 4.}
    \label{fig:Fig8}
\end{figure}

\begin{figure}
    \centering
    \includegraphics[width=\columnwidth]{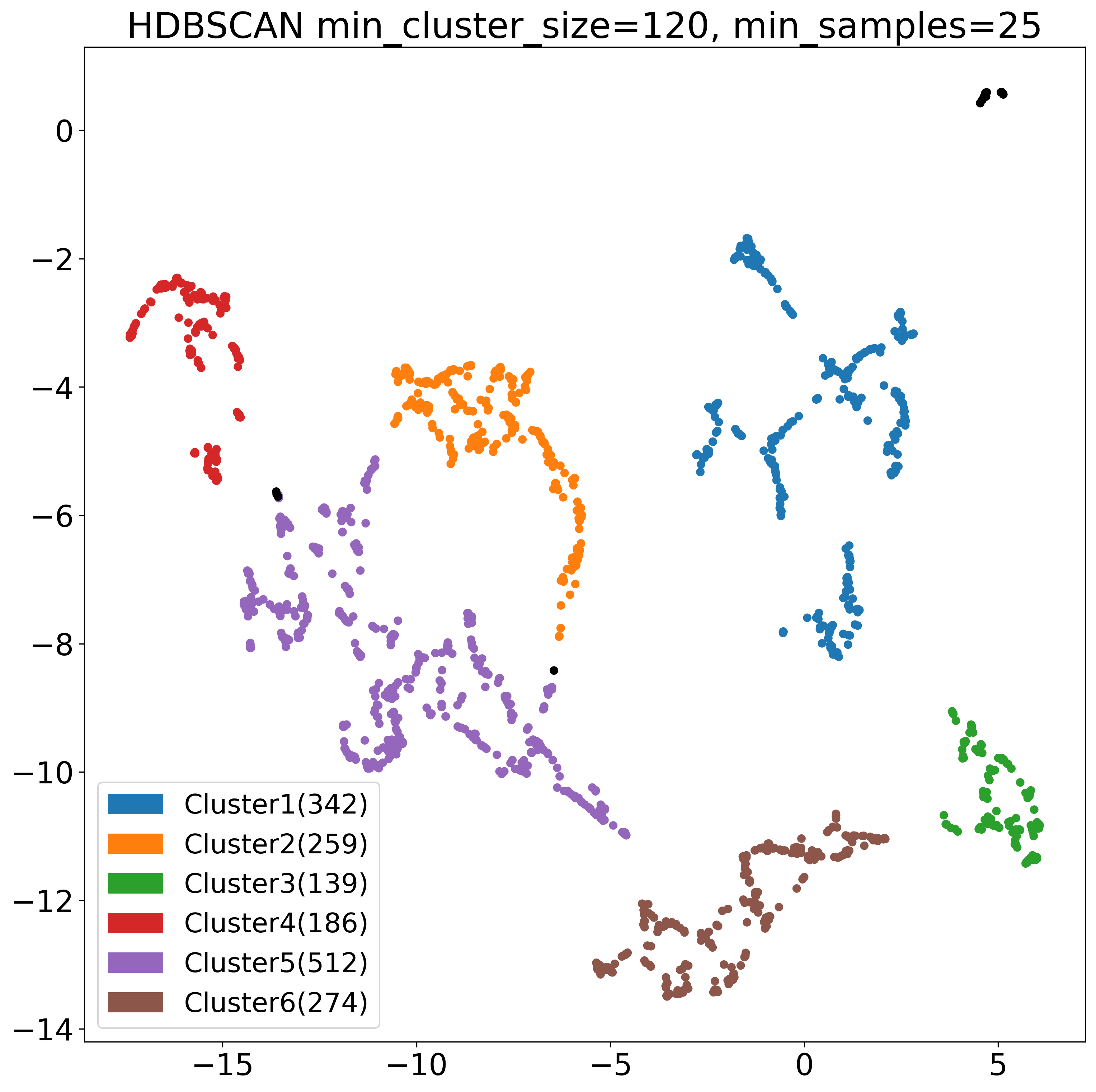}
    \caption{HDBSCAN clustering result with \pyth{min_cluster_size} = 120 and \pyth{min_samples} = 25.}
    \label{fig:Fig9}
\end{figure}

In the main text, we apply the K-means algorithm with \pyth{n_cluster} = 3 to cluster the result provided by the unsupervised machine learning result. Nevertheless, it is not the only way to present the cluster. In this section, we present another result using K-means but set \pyth{n_cluster} = 4, and the other result with HDBSCAN, a clustering algorithm that does not require a predetermined total cluster count. The hyperparameter setting for HDBSCAN is \pyth{min_cluster_size} = 120 and \pyth{min_samples} = 25. The two results are presented in Fig.~\ref{fig:Fig8} and Fig.~\ref{fig:Fig9}, respectively.

In Fig.~\ref{fig:Fig8}, we find that the \pyth{n_cluster} = 4 case does not identify the possible sub-cluster at the lower right. Instead,  it divides the biggest apparently single cluster (cluster 2 in Fig.~\ref{fig:Fig2}) into two. Furthermore, \pyth{n_cluster} = 4 has a worse silhouette score than \pyth{n_cluster=3} in Fig.~\ref{fig:n_cluster}. As a result, we decided to use \pyth{n_cluster = 3} for further analysis in this research.

In Fig.~\ref{fig:Fig9}, we see HDBSCAN is offering a result with significantly more clusters. The cluster count in HDBSCAN is automatically determined, thus we have set \pyth{min_cluster_size} small to identify the sub-clusters. As a result, the possible sub-clusters are mostly identified, at the cost of splitting the largest cluster into several pieces. This cluster result could work as an alternative when an analysis with a more detailed classification of FRB20121102A is needed.

% %%%%%%%%%%%%%%%%%%%%%%%%%%%%%%%%%%%%%%%%%%%%%%%%%%

% Don't change these lines
\bsp	% typesetting comment
\label{lastpage}
\end{document}